SY Stepanov

# PROCESSING HETEROGENEOUS DATA SPACE MEASUREMENT SUBPOLAR TERRITORIES TO FORMULATE STOCHASTIC MODELS ASSESSMENT GEOHAZARDS.

Russian State Hydrometeorological University, Saint-Petersburg.

The Northern Sea Route - a shipping highway running along the northern coast of Russia on the seas of the Arctic Ocean, linking the European and Far East ports, as well as the mouth of the navigable Siberian rivers into a single transport system. Otherwise, the main Russian naval communications in the Arctic. Particularly important are strategic factors related to geopolitical and transnational maritime values in the Arctic zone. The role and importance of the Northern Sea Route as a transport route, is determined primarily by the needs of industrial development and the successful functioning of economic complexes adjacent to the track area of the Arctic coast as an important factor in stabilizing the economy and ensuring national geopolitical and economic security of Russia.

In the foreseeable future, the Northern Sea Route is a key element to ensure the transport of the Arctic regions of Russia. Moreover, in the near future its role may increase substantially.

To select the optimal route of navigation required operational ice information covering the area of climate change and the spread of ice in the Arctic. That is why it is so important to know and predict the ice conditions in the circumpolar area [1,2,3].

Through collaboration with the laboratory of satellite oceanography RSHU it was received a huge amount of material placed on the web-resource http://satin.rshu.ru, required to perform the planned project, whose main purpose is to analyze existing statistical data space measurements and processing of heterogeneous data develop a stochastic model geohazard assessment polar territory.

The study used mathematical methods of comparative analysis, as well as the methods of probability theory and mathematical statistics, as a result were excluded from the study area, some geographic data, not subject to the selection criteria and selected data reflecting the density of sea ice in the polar area.

Due to the fact that the main obstacle to the passage of vessels in the Arctic, as well as to select the optimal route is operational ice information covering the area of climate change and the spread of the ice, the selected information is the underlying cause of forecasting of ice conditions in the polar areas.

Due to the limited access to heterogeneous data online SALW, Geoinformation, reflecting the density of sea ice in the Arctic has been obtained by means of remote access via the programming language python. With the assistance of experts from the laboratory was made recoding data available in the required format through an original script:

https://github.com/SOLab/solab-data-scripts/blob/master/point_data.py.

With this code it was connected to a database of satellite oceanography laboratory RSHU Sort the data by time interval, based on which it was found that the available information on the density of the ice in the Arctic is collected in large numbers with a significant interval of time and a large file size, complicating further analysis of the system. As a result of the revealed information was produced by the search for optimal data for a certain period of time, at certain points in the polar area.

The format of presentation of data at the end of the coding was defined as table - .XLSX (Microsoft Office Excel). The second stage of data for further manipulation of geo-information, was the implementation of a database Ice.accdb. The database has been developed in order to store the input primary density information in the Arctic ice, and continuously monitoring the integrity of data.

In order to visualize and further work on the findings it was implemented software application. The program has an intuitive interface and using queries may reflect the latest information about the density of the ice for a selected period of time during from 01.01.2012 to 24.08.2013 year. Current information on the density of ice collected from four points lying over the Northern Sea Route. Each point represents the area of 25 km2. Dots are arranged in the coordinates (50.80) (135.85) (173.95), (193.132) as shown in Figure 1.

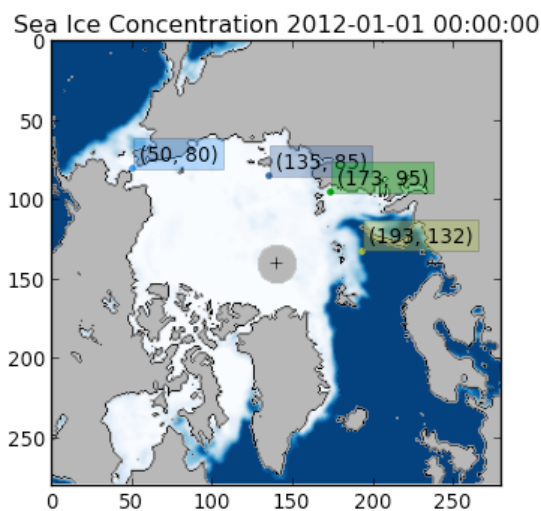

Figure 1 - Arctic grid.

Each point has a specific set of information (see Figure 2):

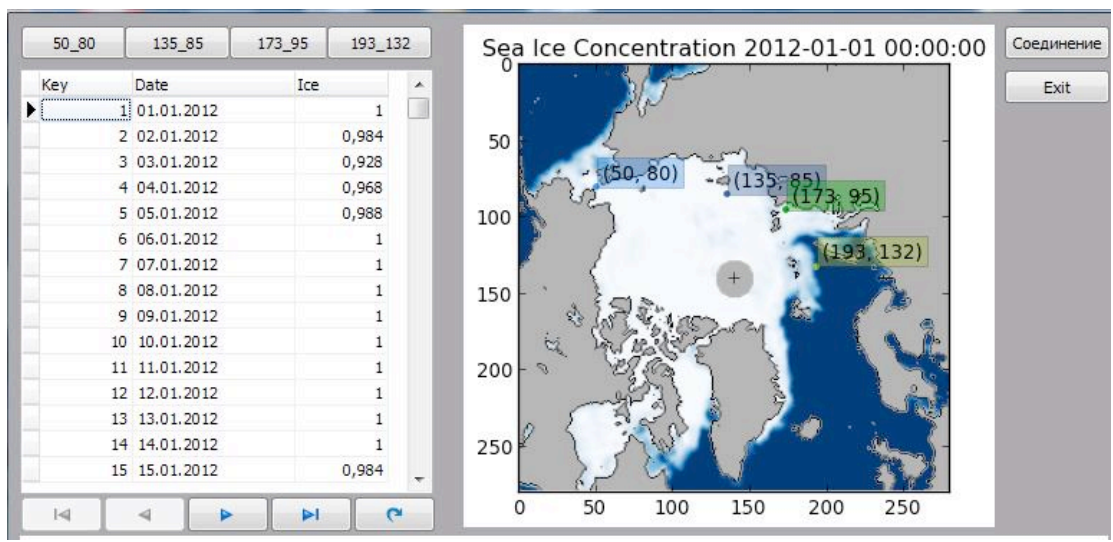

Figure 2 - acquisition time companion, a density of ice.

Time data collection companion - daily at 00:00.

Indicator ice density is in the range from 0 - absent in a given point of the ice, and 1 - in this point the ice density is 100%. Performance will vary depending on the time of year, weather conditions, as well as the selected location on the map.

Initial processing of geo-information can be presented in the following chart (Figure 3):

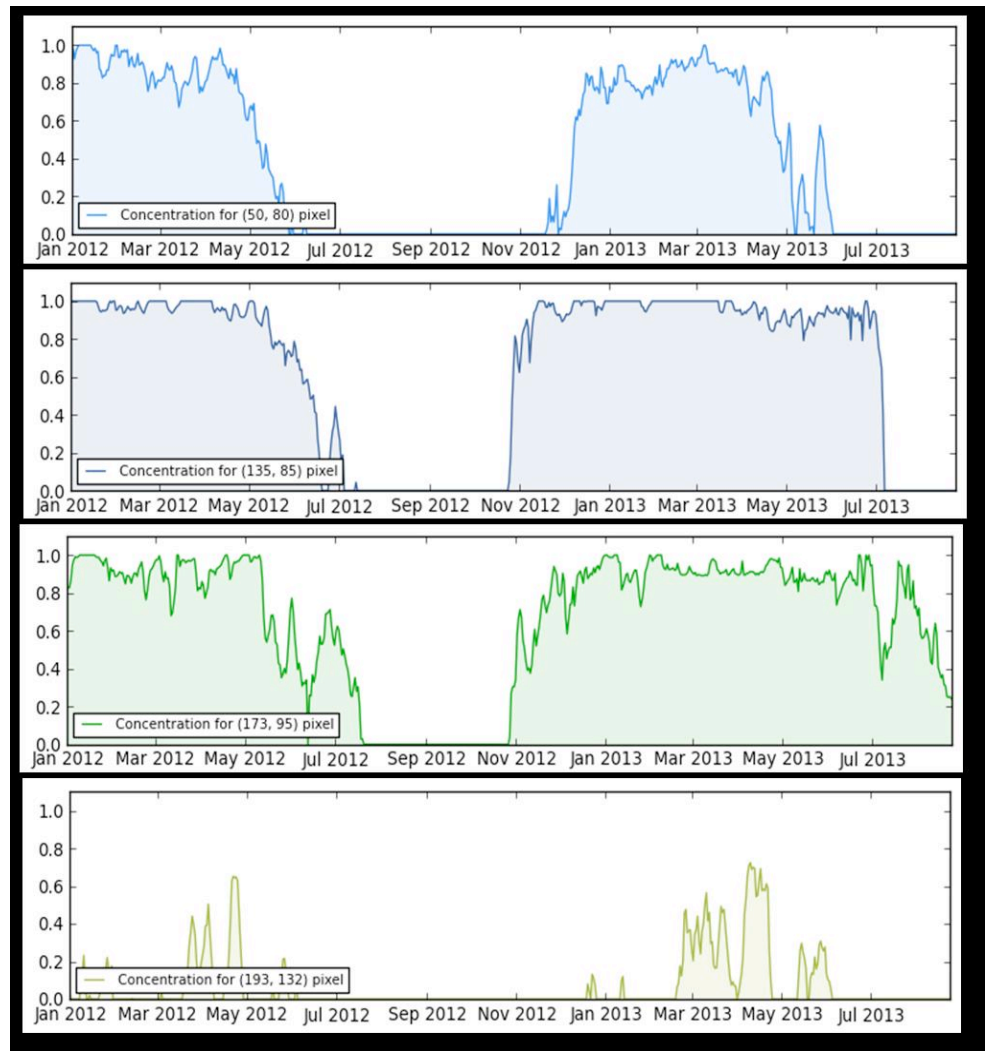

Figure 3 - Graph values depending on the time.

Based on available data, we can determine what points was recorded the greatest density of the ice, and over what period of time. According to the information received was analyzed on the basis of which it was found that the formation of risk assessment model may be used one of the two existing models:
- the use of Kalman filters
- use of risk assessment models in multidimensional geosystems

As a result, research was carried out analysis and classification of statistical measurements of space. It was formed and implemented a program of scientific research. Based on the processed data has been developed a software application for handling various processes with primary information. A classification of geohazards polar areas and formed the requirements for models geohazards.

High information density statistics of ice on the sea surface leads to the possible use of the project results in the prediction of climate change and the spread of the ice edge to select optimal routes of navigation along the Northern Sea Route.

In the future, increasing the number of points, we can determine the most appropriate path (route) for the passage of water transport. After receiving the information for a longer period of time and using the methods of system analysis, it will be possible to predict the densities of ice in the coming season. This will reduce the risk of passing ships on the sea route in the Arctic, and the collapse of getting stuck in sea ice, working on their release, and other rescue operations, will also be debugged system passage flights, increased traffic flow and economic component.